\documentclass[12pt]{iopart}

\usepackage{iopams}

\usepackage{color}

\usepackage{cite}

\usepackage{graphicx, epsfig}

\def \ord{\mathcal{O}}

\def\lba{\left(}    \def\rba{\right)}
\def\lbc{\left[}    \def\rbc{\right]}

\newcommand{\bra}[1]{\langle\left.{#1}\right|}
\newcommand{\ket}[1]{\left|{#1}\right.\rangle}

\begin{document}

\title{Entanglement between particle partitions in itinerant many-particle
  states}






\author{Masudul Haque$^1$, O.~S.~Zozulya$^2$,  and K.~Schoutens$^2$}
\address{$^1$ Max-Planck Institute for the Physics of Complex
Systems, N\"othnitzer Str.~38, 01187 Dresden, Germany}
\address{$^2$ Institute for Theoretical Physics, University of
Amsterdam, Valckenierstraat 65, 1018 XE Amsterdam, the Netherlands}

\begin{abstract}

We review `particle partitioning entanglement' for itinerant many-particle
systems.  This is defined as the entanglement between two subsets of particles
making up the system.
We identify generic features and mechanisms of particle entanglement that are
valid over whole classes of itinerant quantum systems.
We formulate the general structure of particle entanglement in many-fermion
ground states, analogous to the `area law' for the more usually studied
entanglement between spatial regions.
Basic properties of particle entanglement are first elucidated by considering
relatively simple itinerant models.  
We then review particle-partitioning entanglement in quantum states
with more intricate physics, such as anyonic models and quantum Hall
states.

\end{abstract}


%



\section{Introduction}

Bipartite entanglement in many-particle systems, \emph{i.e.}, the
entanglement between one part ($A$) of a system and the rest ($B$),
has grown into a widely studied topic in the last few years.  Usually,
the partitioning is spatial, so that $A$ is a collection of lattice
sites or is a connected region of space.
In this article, we will consider an alternate form of
partitioning, namely \emph{particle partitioning}.
With the wavefunction expressed in first-quantized form, one can meaningfully
partition particles rather than space, and calculate entanglements between
subsets of particles.  Since each particle has a label in first-quantized
wavefunctions, indistinguishability does not preclude well-defined subsets of
particles.  Note that, with such partitioning, $A$ or $B$ do not correspond to
connected regions of space.   Also note  that particle partitioning is
only defined in itinerant systems where the particles hop, and thus has no
meaning for pure spin models.  

Particle partitioning entanglement is generally quite different from the
entanglement between spatial partitions of the same system, and also provides
a distinct set of physical insights compared to the more standard spatial
partitioning entanglement calculations.  Particle entanglement provides a
novel and unique perspective on the structure of itinerant many-particle
wavefunctions.

Given a partitioning, the entanglement can be quantified using various
measures. The basic quantity is the \emph{reduced density matrix} of the $A$
partition, $\rho_A = \tr_B\rho$, obtained by tracing out $B$ degrees of
freedom.  We assume the system to be in a pure state, described by density
matrix $\rho=\ket{\psi}\bra{\psi}$.  Various entanglement quantifiers can then
be extracted from $\rho_A$.  We will mostly confine ourselves to the
entanglement entropy $S_A$, defined as $S_A = - \tr\lbc\rho_A\ln\rho_A\rbc$.
We will also restrict ourselves to zero temperatures, \emph{i.e.}, to 
entanglement in the ground state of itinerant systems.

\emph{Brief history} --- 
Several pieces of work explored simple versions of particle partitioning
entanglement between identical quantum particles, even before the concept was
carefully distinguished from spatial entanglement \cite{Schliemann-etal_PRA01,
  PaskauskasYou_PRA01, Omar-etal_PRA02, Tsinghua_FQHE_PRA02,
  WisemanVaccaro_PRL03, DowlingDohertyWiseman_PRA06,
  IblisdirLatorreOrus_PRL07}.
The relationship between quantum indistinguishability and entanglement was
studied for two fermions in Ref.~\cite{Schliemann-etal_PRA01} and for two
bosons in Ref.~\cite{PaskauskasYou_PRA01}.
Refs.~\cite{Tsinghua_FQHE_PRA02,DowlingDohertyWiseman_PRA06} studied particle
entanglement in somewhat more complicated systems. 

A careful distinction with spatial entanglement, and a comparison between the
two types of partitioning, appears in Refs.~\cite{our-prl-07, our-prb-07}, in
the context of fractional quantum Hall (FQH) states.  
Particle partitioning is tempting in entanglement considerations for FQH
states, because FQH model wavefunctions (\emph{e.g.}, Laughlin states) are
often written in first-quantized form where the particles have explicit
labels.  Thus, Refs.~\cite{Tsinghua_FQHE_PRA02,IblisdirLatorreOrus_PRL07} have also
computed particle-partitioning entanglement in FQH states.
%

In Refs.~\cite{our-prl-07, our-prb-07} and in work reported since then
\cite{SantachiaraCabra_JStatMech07, KatsuraHatsuda_CalogeroSutherland_JPA07,
  our-pra-08}, particle entanglement has been shown to be a promising novel
measure of correlations.
%
%
In fractional quantum Hall states this type of entanglement reveals the
\emph{exclusion statistics} inherent in excitations of such states
\cite{our-prl-07, our-prb-07}.  Similar insight arises from particle
entanglement calculations in the Calogero-Sutherland model
\cite{KatsuraHatsuda_CalogeroSutherland_JPA07}.
For one-dimensional anyon states, particle-partitioning entanglement is found
to be sensitive to the anyon statistics parameter
\cite{SantachiaraCabra_JStatMech07, SantachiaraCalabrese_anyon}.

\emph{This review} --- 
Clearly, entanglement between particles in itinerant systems is a promising
new concept, potentially useful for describing subtle correlations and the
interplay between statistics and interaction effects.  A broad study of the
concept and its utility is obviously necessary.
In this review, we will survey the results that are available until
now.  We will focus in particular on common features and on results of
wide generality, that provide insights into classes of quantum
itinerant systems.

The present review is solely concerned with particle-partitioning entanglement
in itinerant many-particle systems.  We will therefore not discuss
entanglement between spatial partitions, or any other kind of entanglement.
Reviews of other types of entanglement can be found, \emph{e.g.}, in
Refs.~\cite{Amico-eta_RMP08, EisertCramerPlenio_AreaLawReview_arxiv08}.  The
other reviews of this special issue provide more recent and more
condensed-matter-oriented perspectives on entanglement in many-particle
states.

The target audience for this review is condensed matter physicists interested
in various possible kinds of entanglement in many-particle wavefunctions.  As
such, other than using the definition of the entanglement entropy $S_A$, we do
not treat or use any quantum information theory topics.
Ref.~\cite{Horodecki_RMP09} provides a recent review of quantum entanglement
from that perspective.  It is of course not possible to make a review of the
present type completely self-contained, since we cannot introduce in detail
each of the several models and states considered here.  We therefore assume
familiarity with several classes of many-particle states or models.  Only
minimal motivational background is provided for each model.  The topic here is
particle-partitioning entanglement and not the individual models.
We expect that the typical practicing condensed matter theorist will indeed be
familiar with most of the many-particle models and states employed.

We start in Section \ref{sec_elementary-example} by working out in detail an
elementary example of particle partitioning, contrasted to spatial
partitioning of the same quantum state.  The fact that entanglement depends
crucially on the type of partitioning is perhaps not as widely appreciated as
it should be; we hope a detailed example helps clarify the concept of particle
partitioning.  In Section \ref{sec_general} we present some generic results
and intuitions, before moving on to specific systems in the following
sections.

Sections \ref{sec_bosons}, \ref{sec_fermions}, \ref{sec_anyons} consider
respectively bosons, fermions and anyons, and review numerical and analytical
results in order to provide an overview of various mechanisms for
particle-partitioning entanglement.  We then turn to more unusual
many-particle states: section \ref{sec_FQH} reviews results for
fractional quantum Hall states and section \ref{sec_CG} for the
Calogero-Sutherland models.

\section{An elementary example}  \label{sec_elementary-example}

The concept of particle partitioning causes enough confusion to justify using
a very simple example to illustrate in detail the definition and its
difference from spatial partitioning.  Readers comfortable with the concept
may safely skip this section.

We imagine two (spinless) fermions in three sites (or orbitals), which
we label $\alpha$, $\beta$, $\gamma$.  We will use the wavefunction
\[
\ket{\psi} ~=~ \lba a_1 c_{\alpha}^{\dagger} c_{\beta}^{\dagger} + a_2
c_{\beta}^{\dagger} c_{\gamma}^{\dagger} \rba \; \ket{\rm vacuum} 
~=~ a_1\ket{110} + a_2\ket{011}
\; .
\]
For usual spatial partitioning, we can consider for example partition $A$ to
consist of site $\alpha$ only.  Then the reduced basis for $A$ consists of the
$\alpha$ microstates $\ket{0}$ and $\ket{1}=c_{\alpha}^{\dagger}\ket{0}$, and
the reduced density matrix in this basis is 
$\rho_A = \left(
\begin {array}{cc}
|a_2|^2 & 0\\ 0 & |a_1|^2
\end {array}
\right)$.

One could also take $A$ to contain sites $\alpha$ and $\beta$.  Then the
reduced basis for $A$ contains four states, $\ket{00}$,
$\ket{01}=c_{\beta}^{\dagger}\ket{0}$, $\ket{10}=c_{\alpha}^{\dagger}\ket{0}$,
and $\ket{11}=c_{\alpha}^{\dagger}c_{\beta}^{\dagger}\ket{0}$, and in this
basis 
\[
\rho_A = \left(
\begin {array}{cccc}
0 & 0 & 0 & 0\\ 0 & |a_2|^2 & 0 & 0\\ 0 & 0 & 0& 0\\ 0 & 0 & 0 & |a_1|^2
\end {array}  \right)  
\;  .
\]

We next turn to particle partitioning, for which the wavefunction must be
expressed in first-quantized form with explicit anti-symmetrization:
\[
\fl \ket{\psi} ~=~ a_1 \lbc \phi_{\alpha}(1) \phi_{\beta}(2) -
\phi_{\alpha}(2) \phi_{\beta}(1) \rbc /\sqrt{2} ~+~ a_2 \lbc
\phi_{\beta}(1) \phi_{\gamma}(2) - \phi_{\beta}(2) \phi_{\gamma}(1)
\rbc /\sqrt{2}  \; .
\]
The particles (fermions) now have labels, so that we can consider the
entanglement between particle 1 and particle 2.  (Partition $A$ contains
particle 1.)  Since particle 1 can be in any one of the three sites, the
reduced basis for $A$ can be labeled by the site labels, $\ket{\alpha}$,
$\ket{\beta}$, $\ket{\gamma}$.  The reduced density matrix is
\[
\rho_A = \left(
\begin {array}{ccc}
 |a_1|^2/2 & 0 & -a_1a_2^* \\ 0 & 1/2 & 0\\ -a_1^*a_2 & 0 &  |a_2|^2/2 
\end {array}  \right)  
\;  .
\]
From this simple example we already see that particle partitioning
entanglement is utterly different from spatial or site partitioning
entanglement.
A second lesson that emerges from this example is that particle partitioning
entanglement is affected substantially by the (anti-)symmetrization which is
explicit in first quantization.  It is therefore no surprise that this type of
entanglement is especially sensitive to quantum statistics.

\section{General considerations}  \label{sec_general}

Before analyzing specific systems, we present some facts and
conjectures broadly applicable to a variety of itinerant quantum
many-particle states.

\subsection{Bounds} 
A generic itinerant lattice system has $N$ particles
in $L$ sites; we consider bosons or spinless fermions with
$N\leq{L}$.
In every case, a natural upper bound for $S_n$ is provided by the (logarithm
of the) size of the reduced density matrix $\rho_A = \rho_n$, \emph{i.e.}, the
dimensions of the reduced Hilbert space of the $A$ partition.  This size is
$\left( \begin{array}{c}L\\ n\end{array} \right)  = C(L, n)$
for fermions and $C(L-1+n, n)$ for bosons.  The actual rank of $\rho_n$ can be
much smaller due to physical reasons, so that the entanglement entropies
are usually significantly smaller than the upper bounds, as we shall see in
the examples we treat.

In a bosonic system, $S_n$ can vanish, since a Bose condensate
wavefunction is simply a product state of individual boson
wavefunctions, each identical.  For fermions, however,
anti-symmetrization requires the superposition of product states; for
free fermions described by a Slater determinant wavefunction, this
causes $\rho_n$ to have $C(N,n)$ equal eigenvalues.  This provides a
nonzero lower bound for $S_n$ in a fermionic system.  To summarize:
\begin{equation}  \label{eq_bounds_b}
{\rm Bosons:} \qquad 0 ~\leq~ S_n  ~\leq~ \ln 
\left( \begin{array}{c} L-1+n \\ n  \end{array} \right)
\; , 
\end{equation}
\begin{equation} 
\label{eq_bounds_f}
{\rm Fermions:} \qquad \ln 
\left( \begin{array}{c} N \\ n  \end{array} \right)
~\leq~ S_n ~\leq~ \ln 
\left( \begin{array}{c} L \\ n  \end{array} \right)
\; .
\end{equation}

\subsection{Standard form for fermions} 

For large fermion number, $N{\gg}1$, we propose the following widely
applicable form for the entanglement of $n{\ll}N$ fermions with the rest:
\begin{eqnarray} \label{eq_canonical}
S_n (N) ~&=~  \ln C(N,n) ~+~ \alpha_n ~+~ \ord(1/N^{\gamma}) \\ 
&=~ n \ln{N} ~+~ \alpha'_n ~+~  \ord(1/N^{\gamma})   \; ,
\end{eqnarray}
with $\gamma>0$.  This form is suggested by results reported in
Refs.~\cite{our-prl-07,our-prb-07, KatsuraHatsuda_CalogeroSutherland_JPA07,
SantachiaraCabra_JStatMech07, our-pra-08}. 
For example, $\alpha_n = n \ln m$ for the Laughlin state at filling
$\nu=1/m$ \cite{our-prl-07}.
The same standard behavior seems to hold for bosonic systems which lack
macroscopic condensation into a single mode, \emph{e.g.}, bosonic Laughlin
states \cite{our-prb-07}, or hard-core repulsive bosons in one dimension
\cite{Calabrese_personal}.
Note that, for lattice sizes larger than $N$, the generic behavior
(\ref{eq_canonical}) indicates that the entanglement entropy does not saturate
the upper bound (\ref{eq_bounds_b}) or (\ref{eq_bounds_f}) obtained from the
size of the reduced Hilbert space.

Subtle correlation and statistics effects can be contained in the behavior of
the $\ord(1)$ term $\alpha_n$, and sometimes also the $\ord(1/N^{\gamma})$
term.  To get some intuition about how such effects show up in $\alpha_n$, as
we summarize the behavior of $\alpha_n$ for several kinds of states.  For free
fermions, for charge-ordered states of the spinless-fermion model (subsection
\ref{sec_t-V}, also Ref.~\cite{our-pra-08}), and for Laughlin states (section
\ref{sec_FQH} and Refs.~\cite{our-prl-07, our-prb-07}), we have
\[
 \alpha_n({\rm FF}) = 0 \, , \quad \alpha_n({\rm CDW}) = \ln{2} \, , \quad
 \alpha_n({\rm Laughlin}) = n\ln{m}  \, .
\]
We note that states which are intuitively `more nontrivially
correlated' have stronger $n$-dependence in $\alpha_n$.  This
strongly suggests that the $\alpha_n$ function is a measure of
correlations in itinerant fermionic states.  It is natural to
conjecture that the linear-in-$n$ behavior of $\alpha_n$ is symptomatic
of intricately correlated states like quantum Hall states, and
that in generic itinerant states $\alpha_n$ will have sub-linear
dependencies on $n$.

\subsection{Exceptional case of macroscopic degeneracy}

One has to treat with care cases where the single-particle spectrum has a
highly degenerate ground state with degeneracy larger than $N$.  The case of
fractional quantum Hall (FQH) states is one example that we will treat in
detail in Section \ref{sec_FQH}.  For FQH states, the appropriate Landau level
is immensely degenerate and only partially filled.  Another example is a
macroscopically degenerate ``flat band'' that is partially filled.  Flat bands
appear in the band structure of frustrated lattices (kagome, checkerboard,
pyrochlore,...), where one of the energy bands can be dispersion-free and have
the same energy for all momentum; hence the name `flat'.  (See, \emph{e.g.},
Ref.~\cite{Lacroix_physicaB09} for a discussion of itinerant systems in
flat-band situations due to lattice frustration.)

\begin{figure}
\centering
\includegraphics*[width=0.6\textwidth]{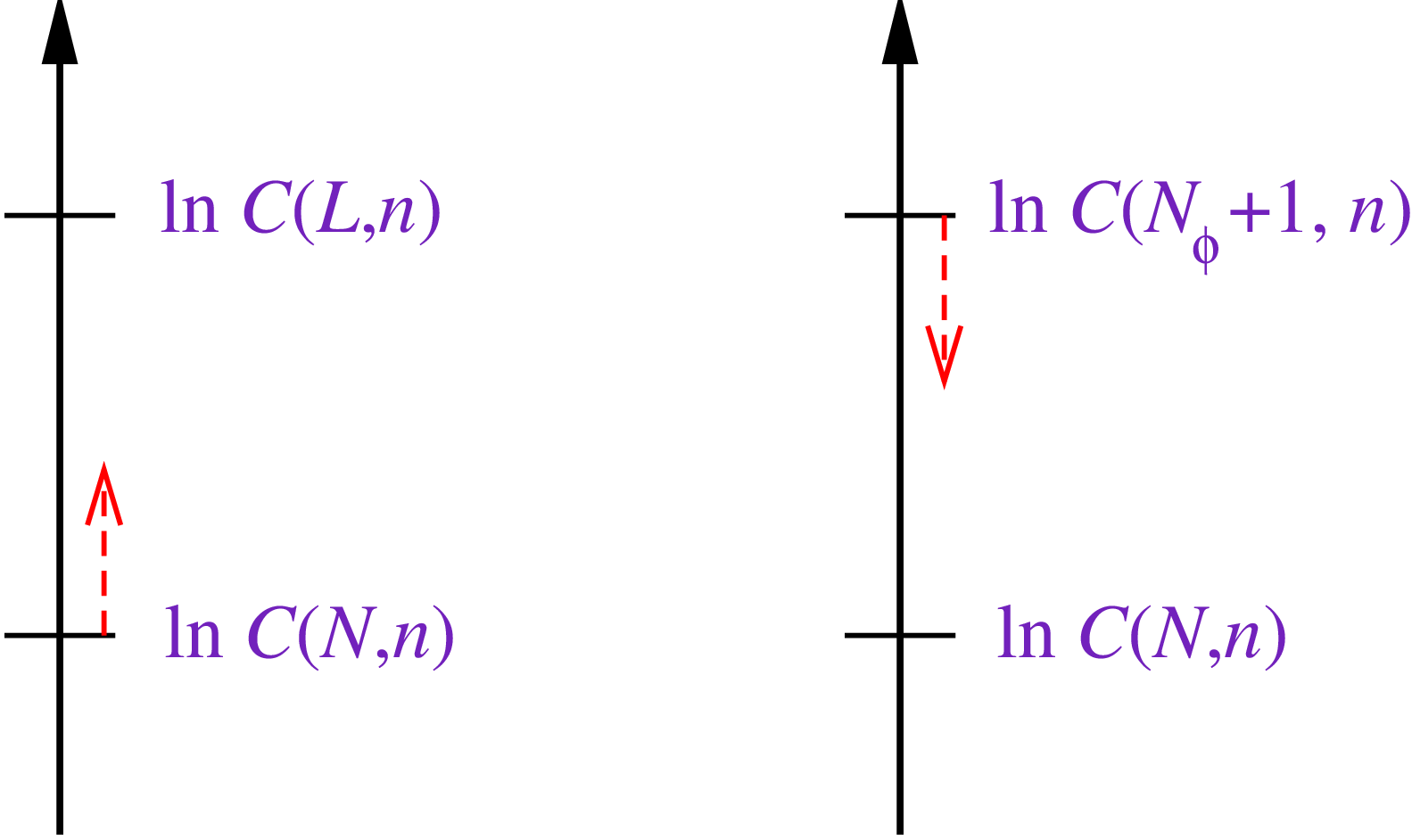}
\caption{ \label{fig_upperlowerbounds}
Lower and upper bounds for particle-partitioning entanglement in fermionic
systems.  Left: usual situation such as mobile fermions in a
non-frustrated lattice.  Right: fermions in a flat band or within a
Landau level, with macroscopic degeneracy.  Dashed arrows show the effect of turning on interactions.
}
\end{figure}

Denoting the degeneracy by $N_{\phi}+1$ (FQH notation), we have the upper
bounds 
\[ 
S_n \leq \ln \left(\begin {array}{c} N_{\phi}+1\\ n \end {array} \right) \; , \quad 
S_n \leq \ln \left(\begin {array}{c} (N_{\phi}+1)+n-1\\ n \end {array} \right) \; ,
\]
respectively for fermions and bosons, from Hilbert-space counting alone.  This
is very similar to the case of $N$ particles in $L$ sites, above.  (The lower
bounds are the same as before.)
The difference is that for non-interacting particles in a flat band,
$S_n$ can be equal to the upper bound, whereas in the usual case it is
equal to the lower bound.  In this case of macroscopic degeneracy,
interactions \emph{reduce} the particle entanglement from the upper
bound, as opposed to the conventional situation where interactions
\emph{raise} the particle entanglement from the lower bound.  This is
illustrated schematically in Figure \ref{fig_upperlowerbounds}.

\subsection{Generic mechanisms: Slater terms; fragmentation or absence of
  condensation; massive degeneracies}

We summarize here the mechanisms through which an itinerant many-particle
system can possess particle partitioning entanglement.  This provides
intuition concerning what physical information is contained in the
entanglement between particle partitions.

An ideal bosonic system is fully condensed in a single mode, and therefore has
$S_n=0$.  Thus, for bosonic systems, particle entanglement is a measure of the
\emph{deviation} from Bose condensation.  Interactions provide a simple
mechanism for this -- since an interacting Bose system is only partially
condensed, it possesses nonzero particle entanglement. Another mechanism is
low dimensionality; hard-core repulsive bosons living on a line (1D continuum;
Lieb-Liniger model) do not condense.  In addition, condensate fragmentation
provides a second mechanism for nonzero $S_n$.  Fragmentation is not as exotic
a phenomenon as commonly perceived; in fact it provides a unifying perspective
to describe Mott phases of bosons  in lattice geometries
\cite{MuellerHoUedaBaym_fragmentation}.  Finally, one can also have $S_n\neq0$
through quantum-mechanical superposition of condensates in different modes,
\emph{i.e.}, Schr\"odinger cat states.  In Section \ref{sec_bosons}, we will
illustrate these situations through several examples.

For fermionic systems, fermionic statistics already provides a
contribution ${\ln}C(N,n)$ to the particle entanglement entropy --
this is the value of $S_n$ when the system wavefunction is a single
Slater determinant.  The excess particle entanglement over this amount
tells us how far one has to go beyond a single Slater determinant in
order to describe the physics of the system.  In other words, the
excess particle entanglement reflects loosely the number of Slater
determinant terms of similar amplitudes that need to be combined to
produce the system wavefunction.

The cases of macroscopic degeneracies require additional considerations.
Particle entanglement can be large here due to the much larger Hilbert space
available without paying an energy cost, so that ground state wavefunctions
can involve much more than a single Slater determinant (fermions) or a few
condensate modes (bosons).

\subsection{Relationship with correlation functions}  \label{sec_corr-fns_general}

The purpose of this subsection is to clarify the relationship between particle
partitioning entanglement and more traditional condensed-matter quantities,
namely, correlation functions. 

In the easiest case, $n=1$, the one-particle entanglement entropy $S_1$ can be
obtained from the one-particle correlation functions, or the single-particle
state occupancies. For example, for one-dimensional translationally invariant
systems, $S_1$ can be expressed through momentum occupation numbers. The
momentum occupation numbers are Fourier transforms of the reduced density
matrix: $c(k) = L^{-1} \int_0^Ldx\rho_1(x)\cos(2\pi{k}x/L)$, with $L$ being
the size of the system. Then
\[
S_1 = - \sum_{k} c(k) \ln c(k)  \; .
\]

More generally, for $n>1$ it is intuitively clear that the elements of the
reduced density matrices are proportional to correlation functions.  We write
down the precise relationship for a continuum one-dimensional case:
\[
\rho_n(\vec{x}_n, \vec{y}_n) 
= \frac{(N-n)!}{N!} \langle \Psi | \phi^{\dag} (x_1 ) \ldots \phi^{\dag}(x_n) \phi(y_n) \ldots \phi(y_1) | \Psi \rangle ~.
\]
where the `vectors' $\vec{x}_n$, $\vec{y}_n$ encode coordinates of $n$
particles, and 
\begin{eqnarray*}
\rho_n(\vec{x}_n, \vec{y}_n) 
=  \int d z_{n+1} \ldots d z_N  & \Psi^*(x_1, \ldots, x_n, z_{n+1}, \ldots, z_N) \\
 &  \times \Psi(y_1, \ldots, y_n, z_{n+1}, \ldots, z_N)  \; .
\end{eqnarray*}
The modification to lattice cases or higher dimensions is obvious.

\section{Bosonic systems: role of condensation}  \label{sec_bosons}

In this section we illustrate the interplay between particle partitioning
entanglement and Bose condensation, through several example models.  First,
considering a two-site Bose-Hubbard model, we demonstrate nonzero
particle-partitioning entanglement through \emph{condensate fragmentation} and
the formation of \emph{Schr\"odinger cat states}.  Next, consideration of a
lattice boson model and the continuum Lieb-Liniger model reveal bosonic
particle entanglement due to partial condensation and absence of condensation,
respectively.

\subsection{Toy model: two-site Bose-Hubbard} 

Following Ref.~\cite{our-pra-08}, we consider $N$ bosons on a two-site
`lattice', subject to a Bose-Hubbard model Hamiltonian.
The Hamiltonian is
\begin{equation}  \label{eq_BoseHubbard-2site}
\hat{H} ~=~ - \lba \hat{b}_1^{\dagger}\hat{b}_2 +
\hat{b}_2^{\dagger}\hat{b}_1 \rba
~+~ {\frac{1}{2}}U \lba \hat{b}_1^{\dagger} \hat{b}_1^{\dagger} \hat{b}_1 \hat{b}_1
+ \hat{b}_2^{\dagger} \hat{b}_2^{\dagger} \hat{b}_2 \hat{b}_2 \rba   \, .
\end{equation}
We label the $N$-boson basis states by site occupancies, \emph{i.e.}, as
$\ket{N_1,N_2}= \ket{N_1,N-N_1}$.

For $U=0$, the system is a non-interacting Bose condensate, with each boson
packed into the single-particle state
$\frac{1}{\sqrt{2}}\lba\ket{1}+\ket{2}\rba$.
In the $U\rightarrow +\infty$ case, the system is a Mott insulator, with half
the particles in site 1 and the other half in site 2, $\ket{N/2,N/2}$.  Such a
state is simple in the `site' basis (second-quantized wavefunction), but
involves symmetrization in the `particle' basis (first-quantized
wavefunction), leading to nonzero particle entanglement entropy.

Finally, the $U\rightarrow -\infty$ limit involves all particles in either
site 1 or site 2.  The ground state is a linear combination of these two
possibilities, $\frac{1}{\sqrt{2}}(\ket{0,N}+ \ket{N,0})$, which for large $N$ is a
macroscopic `Schr\"odinger cat' state.
Such a state is somewhat artificial, because an infinitesimal energy imbalance
between the two states will `collapse' this state.
For example, a `symmetry-breaking' term of the form $\epsilon \;
\hat{b}_1^{\dagger} \hat{b}_1$, added to the Hamiltonian
(\ref{eq_BoseHubbard-2site}), would favor site 2 and destroy the cat state.
The resulting state is a product state with zero particle entanglement.

\begin{figure}
\centering
\includegraphics*[width=0.85\textwidth]{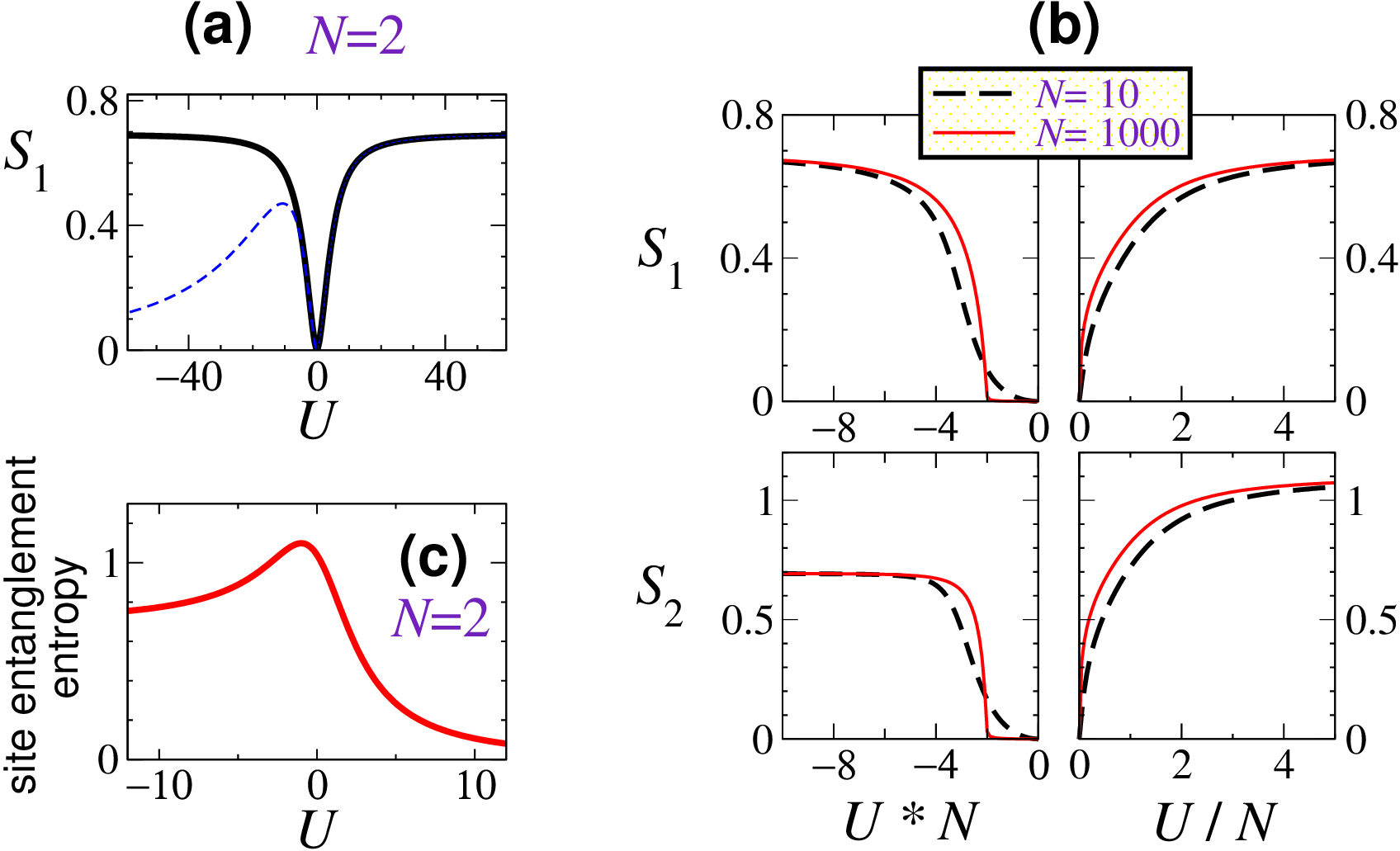}
\caption{ \label{fig_bosons-2sites}
(a,b) Particle partitioning entanglement for bosons in the ground state of a
two-site lattice model with on-site repulsion.
(a) Two bosons.  Solid curve is for the basic Bose-Hubbard model.  Dashed
curve illustrates fragility of `cat' state via a $\epsilon \hat{b}_1^{\dagger}
\hat{b}_1$ term ($\epsilon=0.1$).
(b) 1-particle and 2- particle entanglement entropies for $N=10$ (1000) bosons
in two sites.  Note different units for positive and negative $U$.
(c) Entropy of spatial entanglement between the two sites ($N=2$). 
}
\end{figure}

For the simplest case of two bosons, there is only one way of partitioning
($n=1$), so the only $S_n$ is $S_1$.
We expect $S_1=0$ at $U=0$, and maximal entanglement $S_1=\ln{2}$ for
both `Mott' state at $U={+}\infty$ and the `Schr\"odinger cat' state
at $U={-}\infty$.
The Hilbert space is small; one can diagonalize the problem and calculate
$S_1$ analytically as a function of $U$.  One finds $S_1(U) = S_1(-U)$,
interpolating smoothly between zero and $\ln2 \simeq 0.6931$ in both positive
and negative directions (Figure \ref{fig_bosons-2sites}a).

Figure \ref{fig_bosons-2sites}a also demonstrates the fragility of the cat
state by showing the effect of an $\epsilon \hat{b}_1^{\dagger} \hat{b}_1$
term (dashed curve).  There is no appreciable effect for $U>0$, but for $U<0$
the cat state is destroyed and we get $S_1\rightarrow{0}$ for
$U\rightarrow{-}\infty$.

%
For $N>2$ bosons, it is meaningful to study $S_n$ with $n>1$.  The $n$-particle
reduced Hilbert space has dimension $n+1$; the reduced-space basis states can
be labeled by the number of $A$ bosons in site 1.  In the Mott state
$\ket{N/2,N/2}$, only the diagonal elements of $\rho_n$ are nonzero and they
are all equal; hence $S_n(U\rightarrow\infty)= \ln(n+1)$.  In the cat state,
only two elements are nonzero, both on the diagonal; hence
$S_n(U\rightarrow-\infty)= \ln{2}$, independent of $n$.
Figure \ref{fig_bosons-2sites}b demonstrates, via calculation from wavefunctions
obtained by numerical diagonalization, that $S_n$ increases to $\ln(n+1)$ and
$\ln{2}$ in the $U\rightarrow\pm\infty$ limits.  On the negative side, the
change occurs sharply (around $U=-2/N$); the ground state remains nearly
unentangled between $U=0$ and $U=-2/N$.

Both $\rho_n(U)$ and $S_n(U)$ can be understood in greater detail using
available approximations for the two-site model
\cite{MuellerHoUedaBaym_fragmentation}; a description is given in
Ref.~\cite{our-pra-08}.


To summarize, in the two-site Bose-Hubbard model the Mott state for $U>0$ and Schr\"odinger cat
state for $U<0$ both possess particle entanglement.  The particle entanglement
in the two regimes have different physical origins: the physics of fragmented
condensation for $U>0$ and that of Schr\"odinger cat states for $U<0$.

\emph{Comparison with spatial entanglement}  ---
In Figure \ref{fig_bosons-2sites}c we plot the entropy of entanglement between
the two sites, \emph{i.e.}, the `spatial' entanglement.  The spatial
entanglement is zero in the Mott regime of $U\rightarrow+\infty$, where the
wavefunction is a product state in the second-quantized wavefunction.
(Symmetrization plays no role.)  In the Schr\"odinger cat regime of
$U\rightarrow-\infty$, the spatial entanglement is $\ln2$ like the particle
partitioning entanglement.  In the Bose condensate regime of small $U$, the
spatial entanglement between sites is large, and for $N=2$ peaks at $\ln3$ at
some small negative $U$.  Thus particle partitioning and spatial partitioning
lead to very different entanglements, except for the $U\rightarrow-\infty$
limit.

\subsection{Hard-core bosons on one-dimensional lattice}

Having considered fragmentation and cat states through the two-site model, we
now turn to imperfect or partial condensation.  One way to access such a state
is through the model of hard-core bosons on a 1D lattice (forbidden multiple
occupancy, $U=\infty$) with attractive nearest-neighbor interaction $V$:
\[
H ~=~ - \sum_{i} \lba c^{\dag}_i c_{i+1} +  c^{\dag}_{i+1} c_{i}
\rba ~+~ V \sum_{i} n_i n_{i+1}  ~+~  U \sum_{i} n_i (n_{i}-1)   \; 
\]
with $U\rightarrow\infty$.  We consider $N$ bosons in $L$ sites, subject to
periodic boundary conditions.  This is closely related to the spinless fermion
model treated later in subsection \ref{sec_t-V}.

The point $V=-2$ has a `simple' ground state known exactly \cite{YangYang66}.
This wavefunction is a symmetric equal-amplitude combination of all possible
$C(L,N)$ arrangements of bosons.  
The exact wavefunction can be exploited to yield \cite{our-pra-08}
\[
S_n = \nu n \ln{N} + \ord(N^0)
\]
where $\nu=N/L$ is the filling fraction.  A natural interpretation is that the
pre-factor represents the un-condensed fraction.  Whether this is generic for
bosonic systems with partial condensation remains an open question.

\emph{Details for $n=1$} ---  
The one-particle reduced density matrix is diagonal in the momentum
representation, and has values
\begin{eqnarray*}
\bra{k}\rho_1\ket{k} ~&=~ (N-1)/[L(L-1)]  \qquad & {\rm for}\;\;\; k\neq 0  \\
\bra{k}\rho_1\ket{k} ~&=~  (L-N+1)/L   \qquad & {\rm for}\;\;\; k= 0
\end{eqnarray*}
In the limit $L\rightarrow\infty$ (with fixed filling $\nu=N/L$), the $k=0$
eigenvalue becomes macroscopic at the expense of the others, indicating
off-diagonal long-range order \cite{Yang_ODLRO_RMP62} and partial condensation
with condensate strength $1-\nu$.

\emph{Other cases of imperfect condensation} --- Interacting bosons in three
dimensions also have partial condensation; it would be interesting to
calculate $S_n$ for such a model.  

\subsection{Lieb-Liniger bosons}

The generic one-dimensional continuum boson model with repulsive contact
interactions (Lieb-Liniger model \cite{LiebLiniger_PRB63}) does not possess
Bose condensation, and instead the particles fill up individual-particle
levels just as fermions do.  (In the Bethe ansatz these levels are labeled by
rapidities.)  One can thus expect a leading $S_n\sim{n}\ln{N}$ behavior as for
fermions.  Currently available evidence strongly suggests this to be the
case.  

For $n=1$, the results of Ref.~\cite{SantachiaraCabra_JStatMech07} (reviewed in
Section \ref{sec_anyons}) allow us to infer a leading $\ln{N}$ for the
large-interaction limit, also known as the Tonks-Girardeau limit.  In
addition, unpublished numerical results indicate that the behavior
$S_1\sim\ln{N}$ holds for any nonzero interaction \cite{Calabrese_personal}.

\vspace{1cm}

\section{Fermionic systems: anti-symmetrization and correlations} \label{sec_fermions}

Following Ref.~\cite{our-pra-08}, in \ref{sec_t-V} we use numerical
calculations of the spinless fermion chain with nearest-neighbor interactions
($t$-$V$ model), one of the basic models of correlated-electron physics, to
illustrate particle entanglement in fermionic systems.  Other systems are
commented on in \ref{sec_other-fermion}.

\subsection{Spinless fermions in one dimension}  \label{sec_t-V}

We consider $N$ spinless fermions on an $L$-site chain with periodic boundary
conditions:
\[
H = - \sum_{i} \lba c^{\dag}_i c_{i+1} +  c^{\dag}_{i+1} c_{i}
\rba + V \sum_{i} n_i n_{i+1} \; .
\]
Through a Jordan-Wigner transformation, this model can be mapped to the
anisotropic Heisenberg ($XXZ$) spin chain model with anisotropy parameter
$\Delta=V/2$.
%
%
%

%
For $V=0$ (free fermions), the ground state is simple in terms of
momentum-space modes: a Slater determinant of the $N$ fermions occupying the
$N$ lowest-energy modes.  The $n$-particle reduced density matrix has
${C(N,n)}$ equal eigenvalues, so that $S_n = \ln\lbc{C(N,n)}\rbc$, independent
of the lattice size $L$.

\emph{Half-filling} ---
For repulsive interactions at half filling ($N=\frac{1}{2}{L}$), this model
has a quantum phase transition at $V=2$, from a Luttinger-liquid phase at
small $V$ to a charge density wave (CDW) phase at large $V$.  This mirrors the
well-known transition between gapless $XY$ and gapped Ising phases in the
$XXZ$ model at the Heisenberg point $\Delta=1$ \cite{XXZ-qpt}.

For $N={\frac{1}{2}}L$, the ground state and hence particle entanglement can
be simply understood in the infinite-$V$ limit.
The ground state is an equal superposition of two `crystal' states, and each
of them gives a separate contribution to the reduced density matrix.  The
reduced density matrix has rank $2 C(N,n)$ and equal eigenvalues: $S_n =
\ln\lbc{2C(N,n)}\rbc$.
In the notation of Eq.~(\ref{eq_canonical}), the sub-leading term $\alpha_n$
interpolates between $\alpha_n = 0$ at $V = 0$ and $\alpha_n\rightarrow\ln{2}$
at $V\rightarrow\infty$ for half filling. The interpolation details depend on
$n$ and $N$.

\begin{figure}
\centering
 \includegraphics*[width=0.6\textwidth]{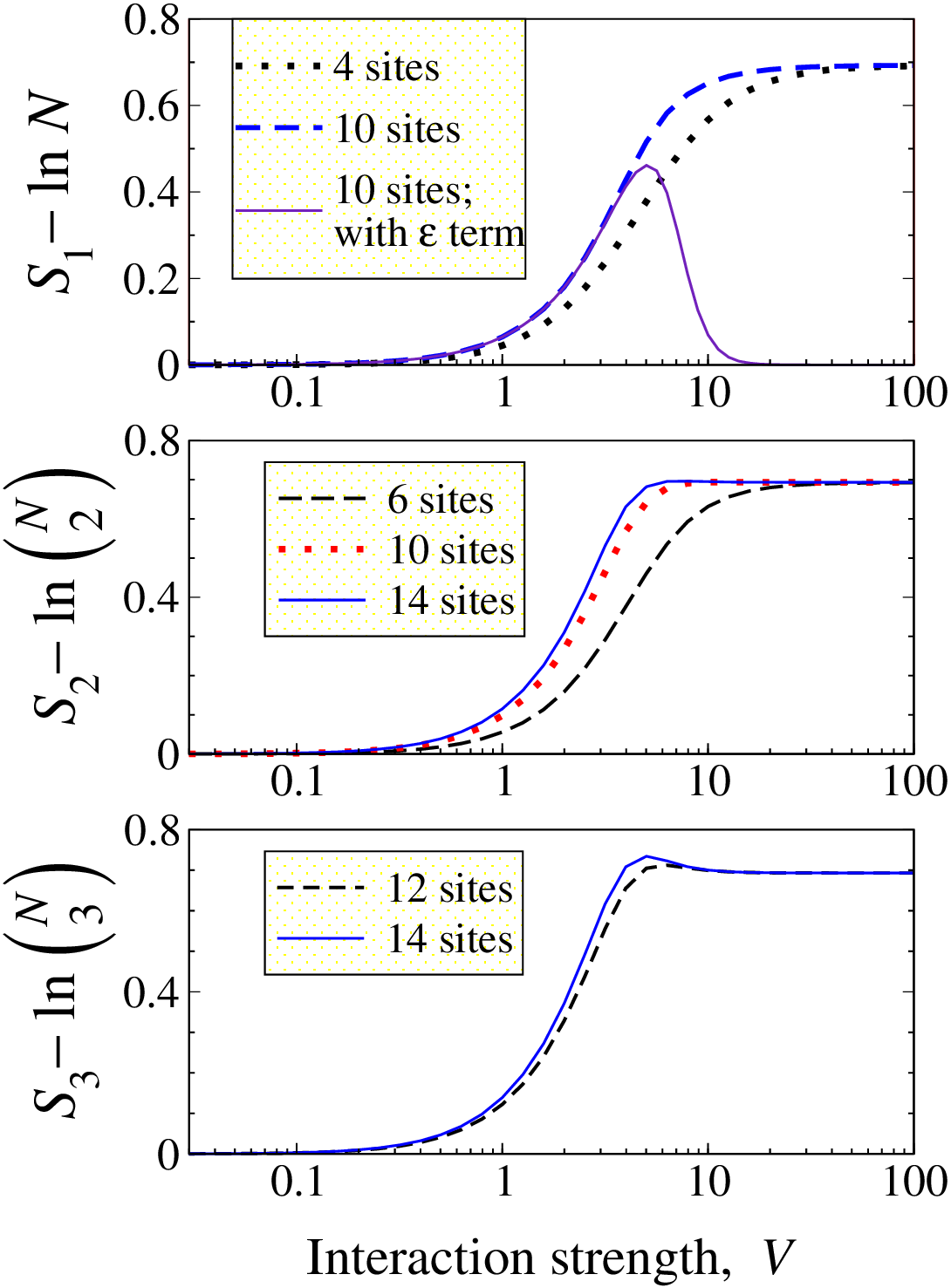}
\caption{ \label{particle-entropies_half-filling}
$n=1$, $n=2$ and $n=3$ entanglement entropy in
  half-filled $t$-$V$ model ($N=L/2$).  The free-fermion contribution
  $\ln\lbc{C(N,n)}\rbc$ has been subtracted off.  The $n=1$ plot also
  displays the effect of a symmetry-breaking
  $\epsilon{c_1^{\dagger}}c_1$ term, with $\epsilon=0.1$.  Inset:
  position of the maximum as function of $\epsilon$.
}
\end{figure}

%
Figure \ref{particle-entropies_half-filling} show $S_n(V)$ for half-filling,
calculated from wavefunctions obtained by direct numerical diagonalization.
The $S_n(V)$ function evolves from $S_{\rm FF}=\ln\lbc{C(N,n)}\rbc$ to
$\ln\lbc{2C(N,n)}\rbc \simeq S_{\rm FF}+0.6931$.
For $n>1$, there is interesting non-monotonic behavior in some cases.
At present there is no simple picture of the non-monotonic behavior.  

We also see Schr\"odinger cat physics in the $t$-$V$ model: the
$V=+\infty$ ground state is a superposition of two CDW states of the
form $\ket{101010...10}$ and $\ket{010101...01}$.  The fragility of
this cat state can be seen by adding a single-site potential,
$\epsilon{c}_1^{\dagger}c_1$, or a staggered potential,
$\epsilon'\sum_{i}c_{2i}^{\dagger}c_{2i}$. The ground state then
collapses to a single crystal wavefunction, and $S_n$ drops to
$\ln\lbc{C(N,n)}\rbc$ (Figure \ref{particle-entropies_half-filling} top
panel).

\emph{Phase transition} ---
The small-$n$ particle entanglement entropies show no strong signature of the
phase transition at $V=2$, even after extrapolating to the
$N\rightarrow\infty$ limit.  
%
%
This is not too surprising because the notion of space enters rather weakly in
the definition of particle entanglement; thus $S_n$ is not too sensitive to
diverging correlation length or large-scale fluctuations.  It remains unclear
whether sharper signatures appear for finite $n/N$ (as opposed to $n{\ll}N$).
%

\begin{figure}
\centering
 \includegraphics*[width=0.8 \textwidth]{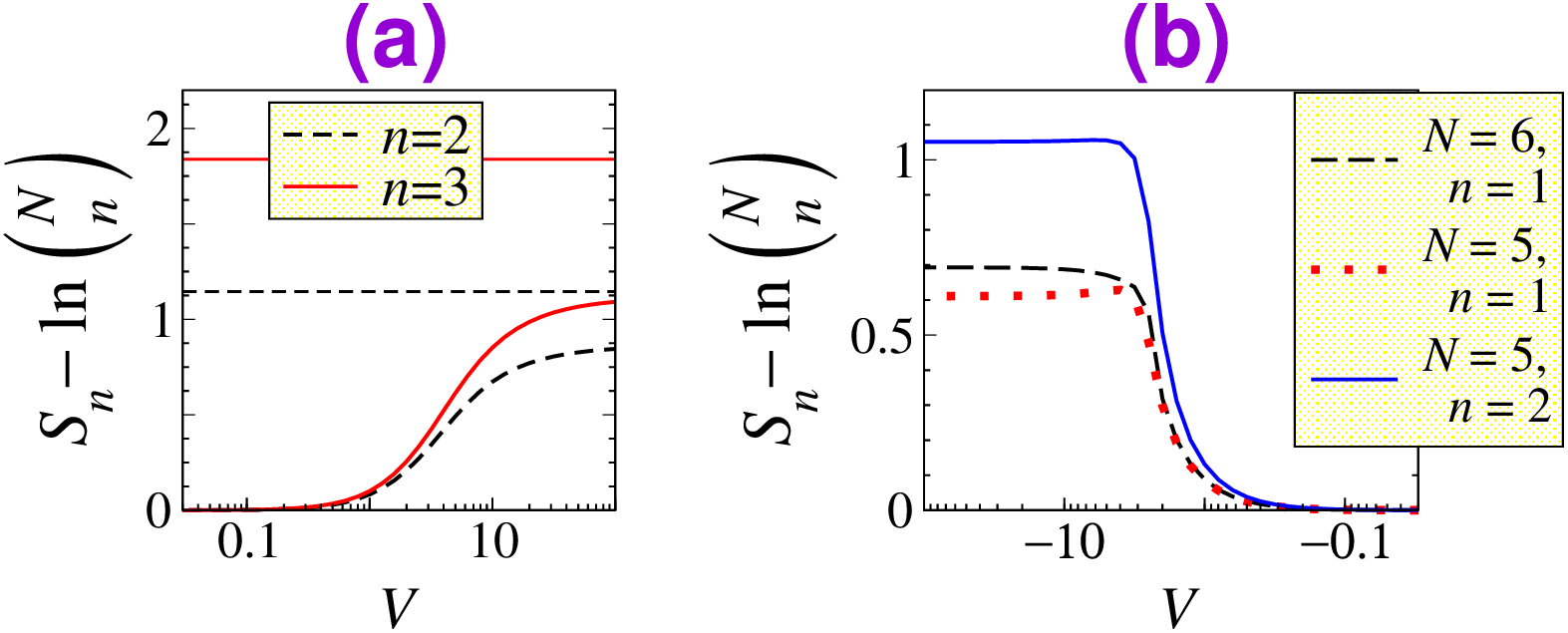}
\caption{ \label{fig_non-hf_negV} 
(a) $S_n$ for $N=7$, $L=12\neq{2N}$. Horizontal lines are
  corresponding maximal bounds $\ln\lbc{C}(L,n)\rbc$.
(b) negative $V$, half-filling.
Free-fermion contribution $\ln\lbc{C}(N,n)\rbc$ has been subtracted off in
each case.
}
\end{figure}

\emph{Away from half-filling} ---
For $N{\neq}L/2$, the behavior is qualitatively similar to the half-filled
case, $\alpha_n$ increasing from zero to an $\ord(1)$ value as $V$ increases
from zero to infinity (Figure \ref{fig_non-hf_negV}a).  However, there is no
simple picture for the $V\rightarrow\infty$ limit.  Also, $\alpha_n(V)$
appears to be monotonic, perhaps because $\alpha_n(V\rightarrow\infty)$ is not
constrained as in the half-filled (CDW) case.

Note that, except for $S_{n=1}$ in the half-filled case, the particle
entanglement never saturates the upper bound, $\ln\lbc{C}(L,n)\rbc$,
dictated by Hilbert space size.

\emph{Negative V} ---
An attractive interaction causes the fermions to cluster.  In the
$V\rightarrow -\infty$ limit, the ground state is a superposition (cat state)
of $L$ terms, each a cluster of the $N$ fermions.  The cat state can be
destroyed as in the positive-$V$ case.  For half-filling with even $N$, the
$V\rightarrow -\infty$ wavefunction yields $S_1=\ln{N}+\ln{2}$ (Figure
\ref{fig_non-hf_negV}b).  There are $\ord(N^{-1})$ corrections for odd
$N=L/2$.

\subsection{Other fermionic systems}  \label{sec_other-fermion}

Other than the one-dimensionless spinless fermion model (\ref{sec_t-V}) and
the quantum Hall states (Section \ref{sec_FQH}), we are not aware of further
explicit calculations for fermionic many-particle systems.
Ref.~\cite{DowlingDohertyWiseman_PRA06} calculates particle entanglement of
the Hubbard dimer (2-site Hubbard model), and the current authors have found
that preliminary numerics on short Hubbard chains show behaviors analogous to
the spinless-fermion chain.

\section{Hard-core anyons in one dimension} \label{sec_anyons}

In one and two dimensions, quantum indistinguishable particles need not
transform under exchange as fermions or bosons: rather, a continuum of possible
intermediate cases connects the boson and fermion cases.  Particles with such
intermediate statistics are known as \emph{anyons} \cite{anyons_various}.  

A Bethe ansatz solution is available for the anyonic many-particle
continuum model with contact interactions, and has recently received renewed
attention \cite{anyon_BetheAnsatz}.   The Hamiltonian is 
\[
 H = - \sum_{i}^N \frac{\partial^2}{\partial x_i^2} + \gamma \sum_{i < j}
 \delta (x_i -x_j) \; , 
\]
and the anyonic statistics imposes the condition 
\[
\Psi^\theta(\cdots, x_i,x_{i+1}, \cdots) ~=~   \exp\lbc i (\theta-\pi) \epsilon
(x_{i+1}-x_i)\rbc \;
 \Psi^{\theta}(\cdots, x_{i+1}, x_i, \cdots) \;,
\]
on the many-body wavefunction.  Here $\epsilon(x) = 0$ (or $1$) if $x>0$
($x<0$)and $\theta$ is the anyonic parameter. For $\theta=0$ or $\theta = \pi$
this Hamiltonian reduces to free fermions or Lieb-Liniger Bose gas
correspondingly. The choice of periodic boundary conditions $\Psi(x_1, \cdots,
x_i + L, \cdots, x_N) = \Psi(x_1, \cdots, x_i, \cdots x_N)$ constrains the
anyonic parameter to be an integer multiple of $2\pi/(N-1)$.  Here $L$ is the
system size.

\begin{figure}
\centering
 \includegraphics*[width=0.75\textwidth]{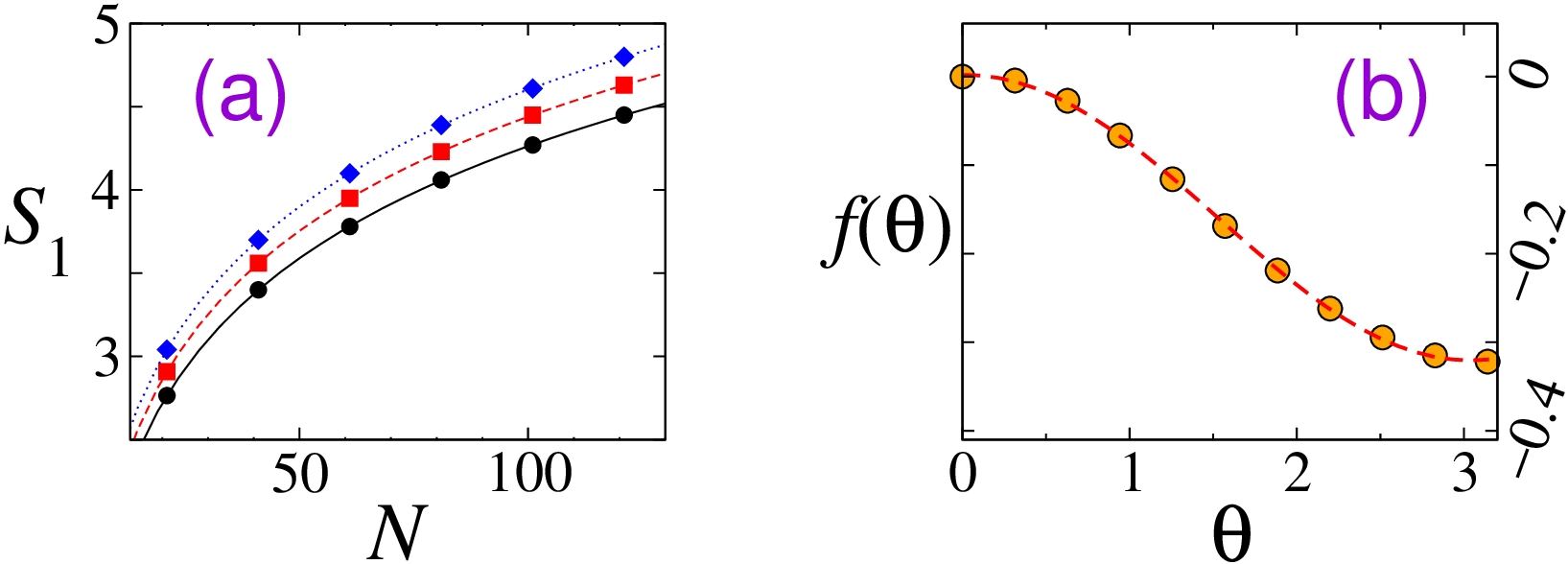}
 \caption{ \label{fig_anyons} 
(a) One-particle entanglement entropy, $S_{1}^{\theta}$, as function of $N$,
   for $\theta=\pi$ (dots fitted by solid line), $\theta=\pi/2$ (squares
   fitted by dashed line), and for $\theta=\pi$ (diamonds fitted by dotted
   line). The fits are of form
   $S_1\sim\ln{N}+f(\theta)+\kappa(\theta)/\sqrt{N}$.
(b) The filled dots are $f(\theta)$  obtained numerically.
The dashed line is a sine fit.
\\ (Data: courtesy of Raoul Santachiara.)
}
\end{figure}

The exact solution of the model,  has been exploited in
Ref.~\cite{SantachiaraCabra_JStatMech07} to calculate the entropy of
particle-partitioning entanglement $S_1$ between $n=1$ anyon and the rest, in
the limit $\gamma\rightarrow\infty$.  In this limit it is possible to compute
the one-particle momentum occupation numbers, $c_N^{\theta}(j) = 1/L \int_0^L
\rho_1(x) \cos(2{\pi}jx/L) dx$.
This in turns allows one to obtain the one-particle entanglement entropy:
\[
S_1^\theta(N) =  ~-~ \sum_{j=-\infty}^{\infty}  c^\theta_N(j) \ln
c^\theta_N(j) ~.
\]
Here $j$ indexes the momenta, $k_j=2\pi{j}/L$.  
For free fermions ($\theta = 0$) there are $N$ equal non-zero momentum
occupation numbers: $c^0_N(j) =1/N $ for $-(N-1)/2\leq j \leq (N-1)/2$. This
gives the expected result $S^{\theta = 0}_1 (N)= \ln{N}$.

For nonzero $\theta$ it is not possible to obtain a closed analytic expression
for the $c^\theta_N(j)$.  Ref.~\cite{SantachiaraCabra_JStatMech07} provides
some asymptotic relations ($j\ll{N}$) using the Toeplitz determinant form of
the one-particle density matrix, but this is not enough to extract the
entanglement entropy.  We will therefore only review numerical results
extracted directly from the exact Toeplitz determinant.
%
%
A fit to this numerical data (Figure \ref{fig_anyons}a) indicates that in the
limit $N \gg 1$ entanglement entropy behaves as:
\[
S_1^{\theta}(N) \simeq \ln N + f(\theta) + \frac{\kappa({\theta})}{\sqrt{N}}
\; .
\]
$f(\theta)$ is an $N$ independent function and describes the main dependence
of the one-particle entanglement entropy on the anyonic parameter $\theta$.
Figure \ref{fig_anyons}(a) shows some $S_1$ values calculated numerically, and
from this one can extract $f(\theta)$ values, plotted in Figure
\ref{fig_anyons}(b).  The extracted $f(\theta)$ values fit extremely well to a
sine function.  Explaining this regular behavior of the sub-leading term
$f(\theta)$ remains an open and important problem.

The fact that entropy scales as $\ln{N}$ with a prefactor $1$ is expected at
$\theta=0$ and $\theta=\pi$ from the arguments involving fermions and
condensation-less bosons, discussed in previous sections.  In this sense, the
leading $\ln{N}$ behavior for intermediate values of $\theta$ is no surprise.
However, from the point of view of the momentum occupation functions
$c^\theta_N(j)$, the prefactor 1 is not evident, as the $c^\theta_N(j)$
functions are quite nontrivial.  
For bosons ($\theta=\pi$), the leading $\ln{N}$ behavior is reported to
persist for finite values of the interaction $\gamma$
\cite{Calabrese_personal}.  One might therefore expect a leading $\ln{N}$ at
all $\theta$ and all $\gamma>0$, but this has not yet been seen through
explicit calculation.

\vspace{1cm}

\section{Fractional quantum Hall states}  \label{sec_FQH}

\subsection{Preliminaries}

The fractional quantum Hall (FQH) states have long fascinated the
condensed-matter community due to their remarkable transport properties and
the exotic nature of their quasiparticle excitations \cite{laughlin-states,
  Moore-Read_NP91, Read-Rezayi_PRB99, Ardonne_Schoutens99}.
%
%
Recently there has been enhanced interest in FQH states with \emph{non-abelian
  statistics} \cite{Moore-Read_NP91, Read-Rezayi_PRB99, Ardonne_Schoutens99},
due to the possibility of implementing quantum computation schemes
topologically protected from decoherence \cite{topological-quantum-computing}.

The unusual features of FQH states have been notoriously difficult to
characterize using traditional condensed-matter concepts such as local order
parameters and $n$-point correlation functions.  Therefore, using novel
measures of quantum correlations, such as entanglement entropies inspired by
quantum information theory, is an attractive idea for quantum Hall states. One
aspect of fractional quantum Hall states, namely \emph{topological order}
\cite{top-order_various}, has been successfully probed using
spatial-partitioning entanglement \cite{KitaevPreskillLevinWen_PRL06,
our-prl-07, our-prb-07, LiHaldane_PRL08, self_entFQHqpt}.
Here, we will describe particle-partitioning entanglement in FQH states,
following mainly Refs.~\cite{our-prl-07, our-prb-07}.
The most striking result is that particle entanglements are closely
approximated by upper bounds whose expressions reflect the \emph{exclusion
  statistics} inherent in FQH states.

We will consider both abelian and non-abelian FQH states, focusing on a
paradigmatic example of each class, namely, the Laughlin (L) states
\cite{laughlin-states} and the Moore-Read (MR) (or pfaffian) states
\cite{Moore-Read_NP91, Read-Rezayi96, Read_2000-overview}.  In planar
geometry, the respective wavefunctions are given by
\begin{eqnarray*}
& \Psi_{\rm L}(\{z_i\}) & =\prod_{i<j} (z_i-z_j)^m e^{- \sum_i |z_i|^2/4}
\nonumber \\
& \Psi_{\rm MR}(\{z_i\}) &= {\rm Pf} \left( \frac{1}{z_i-z_j} \right)
                    \prod_{i<j} (z_i-z_j)^m e^{- \sum_i |z_i|^2/4} \ ,
\end{eqnarray*}
with ${\rm Pf}$ denoting the antisymmetric Pfaffian symbol, and $z_i=x_i+iy_i$
representing the coordinates of the $i$-th particle.  
The fermionic Laughlin states (odd $m$) describe the physics of the most
prominent series of FQH states observed as Hall resistivity plateaus in
transport measurements on two-dimensional electron gases in semiconductor
heterostructures.  The Moore-Read state with $m=2$ is widely expected to
describe the state causing an observed plateau at a Landau-level filling
fraction with even denominator.  Since quantum statistics plays an important
role for particle-partitioning entanglement, we also consider \emph{bosonic}
FQH states.  These have not yet been realized experimentally, but have been
the focus of experimental proposals and efforts with laser-cooled trapped
bosonic atoms.
%

We will describe FQH states in a spherical geometry \cite{Haldane_FQHE_PRL83}.
In this representation the fermions are placed on a sphere containing a
magnetic monopole.  The magnetic orbitals of the relevant Landau level are
then represented as angular momentum orbitals; the total angular momentum is
half the number of flux quanta, $L=\frac{1}{2}N_{\phi}$.  The $N_{\phi}+1$
orbitals are labeled either $l=0$ to $N_{\phi}$ or $L_z=-L$ to $+L$.  For $N$
particles at fractional filling $\nu=1/m$, one finds the interesting FQH
states for $N_{\phi}=mN-S$, where $S$ is a finite-size shift.  The Laughlin
states appear at $S=m$ while for the Moore-Read states $S=m+1$.  The
``filling'' acquires the usual meaning $\nu =N/N_{\phi}$ only in the
thermodynamic limit.
The orbitals are each localized around a ``circle of latitude'' on
the sphere, with the $l=0$ orbital localized near one ``pole.''

\subsection{Summary of main results}

Appreciating that derivations involving FQH states are not readily
accessible to readers unfamiliar with the quantum Hall literature, we
summarize our main results in this subsection.  The technical
details are deferred to the remaining subsections.

We consider the entropy of entanglement between $n_A$ particles of the state
and the remaining $n_B=N-n_A$ particles.  (In this section, we prefer to
display the subscript $A$ explicitly, because there is a profusion of symbols
to distinguish from.)

For both Laughlin and Moore-Read series of states, one can consider how the
structure of the FQH wavefunctions reduce the rank of the reduced density
matrices $\rho_{n_A}$.  Hence one can derive upper bounds $S_A^{\rm bound}$
for the particle entanglement entropy $S_{n_A}$ \cite{our-prl-07, our-prb-07},
which are reduced compared to the naive bound ($S^{\rm F}$) obtained from the
full reduced Hilbert space.  For FQH states on the sphere, in the simpler
cases ($n_A=2$) the rank reduction has a physical interpretation in terms of
the $SU(2)$ multiplet structure of the spectrum of $\rho_{n_A}$.

\begin{figure}
\centering
 \includegraphics*[width=0.65\textwidth]{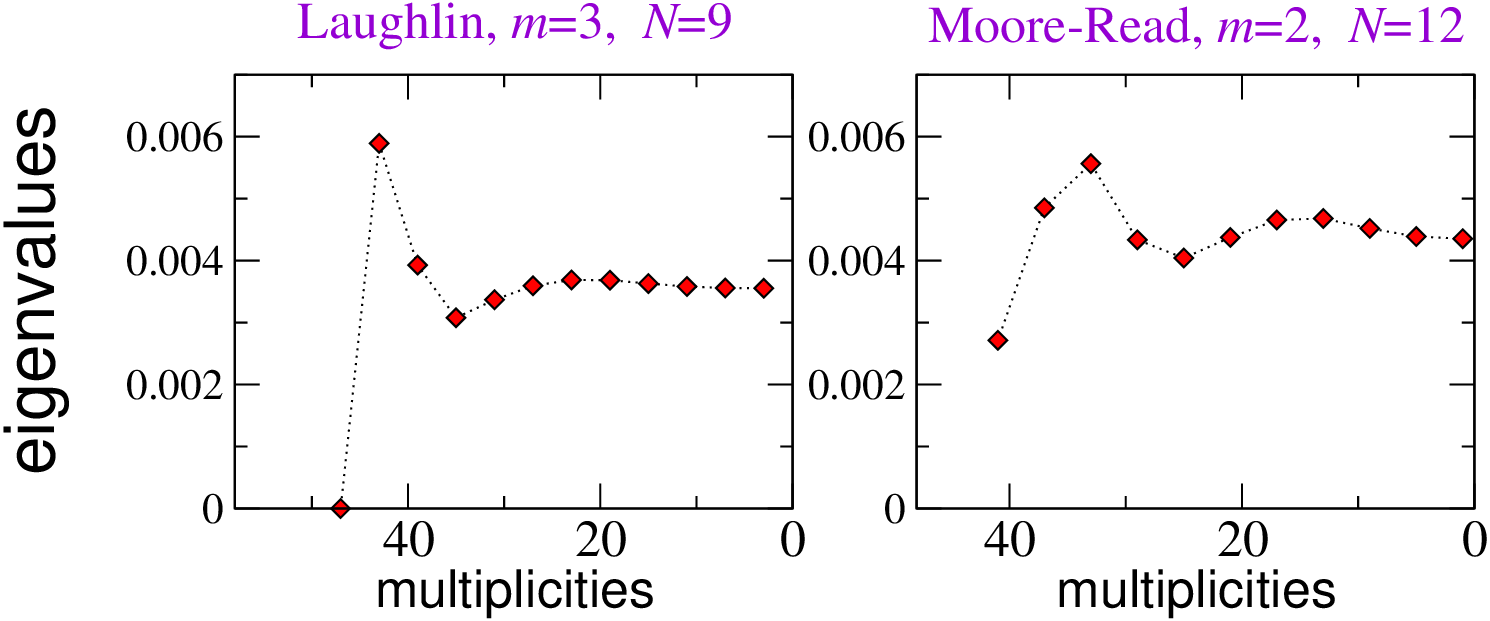}
\caption{  \label{fig_2-pcle_eigen}
Eigenvalues for the 2-particle reduced density matrix, plotted against 
their multiplicities, for one Laughlin and one Moore-Read state.
}
\end{figure}

For $n_{\rm A}\ll{N}$, these upper bounds in fact happen to be close
approximations to the actual values.  This is due to the fact that the nonzero
eigenvalues of $\rho_{n_A}$ are distributed relatively flatly (Figure
\ref{fig_2-pcle_eigen}).  (The more flat the eigenvalue distribution is, the
closer the entanglement entropy is to the upper bound $\ln\mathcal{D}$
dictated by the local Hilbert space dimension $\mathcal{D}$.)
This is in sharp contrast to the exponential-like eigenvalue distributions
well-known from spatial entanglement \cite{our-prb-07, CalabreseLefevre_PRA08,
ChungPeschel_PRB00, PeschelKaulkeLegeza_AnnPhys99}.

The upper bounds are logarithms of combinatorial quantities which reflect the
\emph{exclusion statistics} of quasiparticle excitations in FQH states
\cite{Haldane-exclusion_PRL91}.
In addition, these quantities distinguish between the physics of the Laughlin
and the Moore-Read states -- the fact that the leading correlations have a
2-body nature in the $m=3$ Laughlin states and a 3-body nature in the $m=2$
Moore-Read states, is reflected in the $1/N$ expansions of the approximations
$S_A^{\rm bound}$.  


\subsection{Reduced ranks and entanglement upper bounds; fermionic states}

For $N$ fermions, $n_A$ particles in the $A$ block, and the total
number of orbitals given by $N_\phi+1 = 2L+1$, the 
obvious upper limit $S^{\rm F}_A$ from Hilbert space counting is:
\begin{equation}
S_A \leq S^{\rm F}_A = \ln \left( \begin{array}{c} N_\phi+1 \\ n_A
\end{array} \right) \ .
\label{eq_NaiveBound}
\end{equation}
In the FQH states the correlations are such that the particles avoid each
other and the entropy is further reduced. To obtain a handle on this, one may
reason as follows. The model FQH states in the Laughlin and Moore-Read series
can be characterized as zero-energy eigenstates of a Hamiltonian penalizing
pairs and/or triplets of particles coming to the same position. After tracing
out the coordinates for the $B$ set, the dependence on those in the $A$ set is
such that one still has a zero-energy eigenstate. However, the number of
orbitals available to the $A$ particles is larger than what is needed to make
the model FQH state in the $A$ sector, and one instead has a certain number of
quasi-holes on top of the $A$ set model state. The total ground state
degeneracy for this situation has been studied in the literature: see
Ref.~\cite{Read-Rezayi96} for the Laughlin and Moore-Read states and
Ref.~\cite{NAcounting} for the Read-Rezayi \cite{Read-Rezayi_PRB99} and
non-abelian spin singlet (NASS) \cite{Ardonne_Schoutens99} series of
non-abelian FQH states.

\emph{Laughlin states} ---
The $N$-particle Laughlin state is realized on a total of $N_\phi+1$ Landau
orbitals, corresponding to $N_\phi=m(N-1)$ flux quanta. The Laughlin state for
$n_A$ particles would need $N_\phi^A=m(n_A-1)$ flux quanta; we thus have an
excess flux of $\Delta N_\phi= N_\phi-N_\phi^A=m(N-n_A)$. 
%
This corresponds to the presence of $n_{\rm qh}= \Delta N_\phi$ quasi-holes
over the ground state.  According to Ref.~\cite{Read-Rezayi96} each of the
quasi-holes has a number of $n_A+1$ effective orbitals to choose from, with
bosonic counting rules (meaning that two or more quasi-holes can be in the
same effective orbital). This gives a number of quasi-hole states equal to
\[
\fl
\left( \begin{array}{c} (n_A+1) + n_{\rm qh} - 1 \\ n_{\rm qh}       
         \end{array} \right) 
~=~
\left( \begin{array}{c} n_A + n_{\rm qh}  \\ n_{A}       
         \end{array} \right) 
~=~
\left( \begin{array}{c} N_\phi+1-(m-1)(n_A-1)  \\ n_{A}       
         \end{array} \right) 
\ ,
\]
leading to the following upper bound to the entropy $S_A$
\begin{equation}
S^{\rm bound}_A = \ln 
\left( \begin{array}{c}
N_\phi+1-(m-1)(n_A-1) \\ n_A 
\end{array} \right) \ .
\label{eq_LaughlinBound}
\end{equation}
We remark that this expression has a clear interpretation in terms of
exclusion statistics \cite{Haldane-exclusion_PRL91}: the counting factor in
Eq.~(\ref{eq_LaughlinBound}) gives the number of ways $n_A$ particles can be
placed in $N_\phi+1$ orbitals, in such a way that a particle placed in a given
orbital $l$ excludes particles from orbitals $l'$ with $|l-l'|<m$.



\emph{Moore-Read} ---
For the fermionic Moore-Read states at $\nu=1/m$,
with $m=2,4,\ldots$, we can reason in a similar way,
with now $N_\phi=m(N-1)-1$.
As for the Laughlin states we have an excess flux of 
$\Delta N_\phi = N_\phi-N_\phi^A=m(N-n_A)$
but now the number of quasi-holes is twice this number due to
the fact that the fundamental quasi-holes correspond to half
a flux quantum. Thus, $n_{\rm qh}=2\Delta N_\phi$. We now take 
from Ref.~\cite{Read-Rezayi96} the following result for 
the total quasi-hole degeneracy
\begin{equation}
 \sum_{F \equiv n_A {\rm mod}\ 2}^{n_A}
     \left( \begin{array}{c} n_{\rm qh}/2 \\ F \end{array} \right)
     \left( \begin{array}{c} (n_A -F)/2+ n_{\rm qh}
                           \\ n_{\rm qh} \end{array} \right) \ .
\label{eq_MRBound}
\end{equation}
The logarithm of this expression gives us an upper bound $S^{\rm bound}_A$ as
before.

\emph{Expansion in $N^{-1}$} ---
For the Laughlin states, for $n_A\ll N$ we get from equation
(\ref{eq_LaughlinBound}) for large $N$
\[
S^{\rm F}_A - S^{\rm bound}_A  ~=~
 \frac{1}{N} \frac{m-1}{m} n_A (n_A-1)
 ~+~ \mathcal{O}(1/N^2)
\]
%
%
We compare this to the $m=2$ Moore-Read states:
\[
S^{\rm F}_A - S^{\rm bound}_A = 
\frac{1}{N^2} \frac{3}{4} n_A (n_A-1)(n_A-2) + \ldots
\]
The leading deviation from $S_A^F$ is a 3-body term at order $1/N^2$.  This
result nicely illustrates the fact that the leading correlations in the $m=2$
Moore-Read state have a 3-body character: the wave-function vanishes if at
least three particles come to the same position.

For $m\neq 2$ the leading correlations do have a 2-body character, as 
for the Laughlin states: 
\[
S^{\rm F}_A - S^{\rm bound}_A = 
\frac{1}{N} \frac{m-2}{m} n_A (n_A-1) + \ldots
\]



\emph{Other fermionic FQH sequences} ---
The quasi-hole counting rules for the order-$k$ clustered spin-polarized
(Read-Rezayi \cite{Read-Rezayi_PRB99}) and spin-singlet (NASS
\cite{Ardonne_Schoutens99}) states are all known in the literature
\cite{NAcounting}.  They can be used to generalize the upper bounds on
particle entanglement entropy given in this subsection to these more intricate
non-abelian FQH states.

\emph{Bosonic quantum Hall states} ---
We consider bosonic Laughlin states at filling fraction $\nu=\frac{1}{m}$ with
$m=2,4,\ldots$. The naive upper bound to the entropy associated to placing
$n_A$ bosons in $N_\phi+1$ orbitals is
\begin{equation*}
S^{\rm B}_A = \ln \left( \begin{array}{c} 
    N_\phi + n_A\\ n_A \end{array}
                   \right)
\end{equation*}
The expression for $S^{\rm bound}_A$ remains unchanged from the fermionic
Laughlin case, giving the following leading correction in a $1/N$ expansion
\begin{equation}
S^{\rm B}_A - S^{\rm bound}_A = \frac{1}{N} n_A (n_A-1) + \ldots
\label{eq_bLaughlin1overN}
\end{equation}

For a bosonic Moore-Read state, with filling fraction
$\nu=1/m$ with $m=1,3,\ldots$, the leading $1/N$ correction
becomes
\begin{equation}
S^{\rm B}_A - S^{\rm bound}_A = \frac{1}{N} \frac{m-1}{m} n_A (n_A-1) + \ldots
\label{eq_bMR1overN}
\end{equation}
In the case $m=1$ the leading correlations have 3-body character, leading to
the vanishing of the leading $1/N$ correction.  (Similar to the fermionic
$m=2$ MR states.)

\emph{Multiplet picture for rank reduction} ---
For $n_A=2$, we can get a simple picture of the reduction of the entanglement
entropy (or of the rank of reduced density matrix) compared to the naive
bound, through consideration of multiplet structures present in the
eigenspectrum of $\rho_{n_A}$.
For FQH states on a sphere, the $n_A$-particle reduced density matrices
$\rho_{n_A}$ commute with the total angular momentum operators ${\bf
  L}^2_{n_A}$ and $L^z_{n_A}$ of the selected $n_A$ particles.  As a result,
the eigenvalues of $\rho_{n_A}$ are organized in $SU(2)$ multiplet structures:
an eigenvalue for total angular momentum $L_{n_A}$ will be $( 2 L_{n_A} + 1
)$-fold degenerate.

For $n_A=2$ fermions, each having angular momentum $L=\frac{1}{2}N_{\phi}$,
the 2-particle states have total angular momenta $L_2=2L-1$, $2L-3$, $\ldots$,
$1\, (0)$, for $L$ integer (half-integer), giving a total number of
$(2L+1)(2L)/2$ states.  This corresponds to the naive upper bound to the
entanglement entropy:
\begin{equation}
S_{n_A=2} \leq \ln \left[ (2L+1)(2L)/2 \right] = 
\ln\left( \begin{array}{c}  N_\phi + n_A\\ n_A \end{array}\right)  .
\end{equation}
Inspecting the explicit structure of the fermionic Laughlin states
with $m=3,5,\ldots$, one finds that the eigenvalues corresponding to
2-particle states with $L_2=2L-1$, $2L-3$, $\ldots$, $2L-(m-2)$ all 
vanish. The reason is that the correlations in the Laughlin states are 
such that particles cannot come too close together. For example, if a 
first fermion occupies the $l=0$ orbital, localized near the north pole, 
the Laughlin wavefunction has zero amplitude for finding a second 
fermion in orbitals $l=1$, $l=2$, $\ldots$, $l=m-1$. The highest 
possible value of the angular momentum of the two fermions combined is 
thus $L_2=L+(L-m)$. The remaining number of non-zero eigenvalues is 
$(2L+(2-m))(2L+(1-m))/2$, leading to an improved bound on the entropy 
$S_{n_A=2}$ 
\begin{equation}
S_{n_A=2} \leq \ln \left[ (2L+(2-m))(2L+(1-m))/2 \right] 
\end{equation}
which is clearly the $n_A=2$ case of equation (\ref{eq_LaughlinBound}).

For $n_A>2$, the multiplet structures are more complicated, and it is
difficult to generalize the above argument, but one expects that the arguments
of the previous subsection, relying on the quasihole degeneracy, are
equivalent to the vanishing of one or more eigenvalue multiplets.

\subsection{Numerical results}

\emph{Reduced spectra} ---
In deriving the upper bound $S_A^{\rm bound}$ we relied on the fact that a
certain number of eigenvalues of the reduced density matrix vanish. The bounds
would be exact if all non-zero eigenvalues were equal, but there is some
eigenvalue spread,  the bounds overestimate the actual values for the
entropies.

Figure \ref{fig_2-pcle_eigen} plots the eigenvalues for the $n_A=2$ reduced
density matrix for $N=9$ particles on a sphere in the $m=3$ Laughlin state
($L=12$).  The horizontal axis represents the degeneracy $2 L_2 + 1$ of the
eigenvalues, in descending order. The eigenvalue at $L_2=2L-1=23$, with
degeneracy 47, vanishes; the non-zero eigenvalues show some scatter around an
asymptotic value. Due to this scatter the entropy $S=5.509$ is somewhat lower
than the upper bound $S_A^{\rm bound}= 5.533$.

An important difference between the $m=3$ Laughlin and the $m=2$ Moore-Read
states is the absence of vanishing eigenvalues for the 2-particle reduced
density matrix. The eigenvalue distribution shown in
Figure \ref{fig_2-pcle_eigen} (right) illustrates this point.

In the $m=2$ Moore-Read state, there are vanishing eigenvalues
in the reduced density matrix of $n_A\geq 3$ particles. The number of 
nonzero eigenvalues predicted by Eq.~(\ref{eq_MRBound}) agrees with 
numerical results. For example, for $n_A=3$ and $N=10$ particles there 
are 770 nonvanishing eigenvalues, in agreement with Eq.~(\ref{eq_MRBound}).

While the nonzero eigenvalues are not all equal, their distributions are quite
flat, in sharp contrast to the near-exponential eigenvalue distributions for
spatial partitioning.

\begin{figure}
\centering
 \includegraphics*[width=0.8\textwidth]{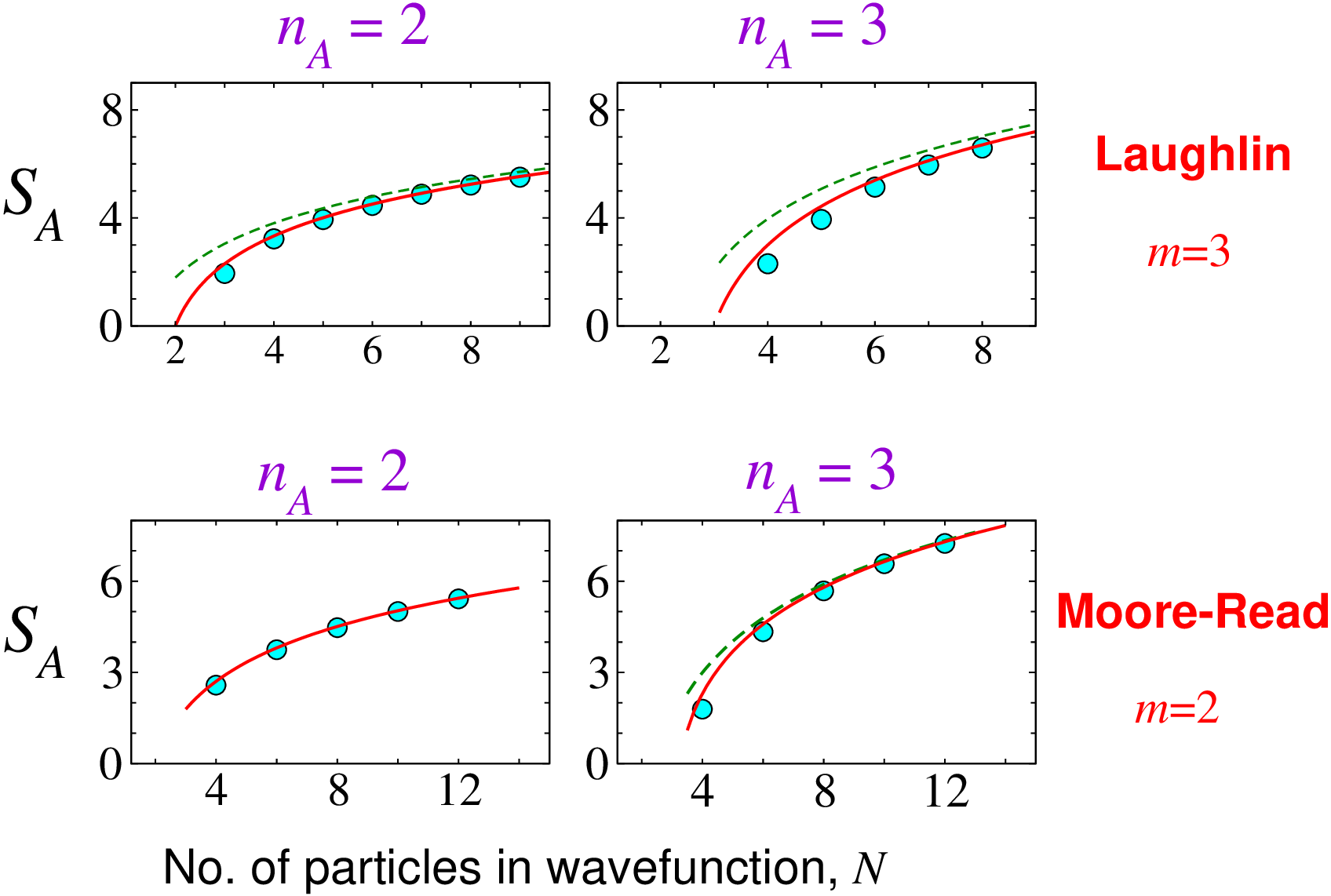}
\caption{  \label{fig_pcle_lghln_MR_23}
Particle entanglement entropies.  Laughlin ($m=3$) state in upper panels,
Moore-Read ($m=2$) state in lower panels.  Dots are numerical exact values,
the dotted line represents $S_A^F$ (eq.~\ref{eq_NaiveBound}) and the solid
curves are the bound $S_A^{\rm bound}$ (eq.~\ref{eq_LaughlinBound} for upper
panels and eq.~\ref{eq_MRBound} for lower panels). }
\end{figure}

\emph{Entanglement entropy values} ---
In Figure \ref{fig_pcle_lghln_MR_23} we compare numerically computed particle
entanglement entropies with the bounds described above.

Note that for the $m=2$ Moore-Read case, $S^{\rm bound}_A$ coincides with
$S_A^{\rm F}$ for $n_A=2$ (dashed and solid lines coincide in lower left panel
of Figure \ref{fig_pcle_lghln_MR_23}), which can be inferred from equation
(\ref{eq_MRBound}) and is also reflected in the fact that none of the
eigenvalue multiplets vanish in that particular case (Figure
\ref{fig_2-pcle_eigen} right panel).

\subsection{Reduced density matrices and correlation functions}
\label{sec_pcle-RDM-n-corrln-fn}

Since the $n_A$-particle reduced density matrices $\rho_{n_A}$ are obtained by
integrating out all but $n_A$ of the particles, one expects these matrices to
be related to the $n_A$-particle correlation functions.  For $n_A=2$,
Ref.~\cite{our-prb-07} provides a detailed explanation of this relationship,
more specific than the general discussion of subsection
\ref{sec_corr-fns_general}.

In particular, the $n_A=2$ eigenvalue distributions are indeed very closely
related to the two-particle correlation function $g_2(r)$.  In fact, the
$g_2(r)$ curve may be regarded as simply a continuous form of the $\lambda_l$
versus descending-$(2l+1)$ curves of Figure \ref{fig_2-pcle_eigen}.  This
explains why the eigenvalue distributions of $\rho_{n_A=2}$ in Figures
\ref{fig_2-pcle_eigen}, although discrete, are reminiscent of the well-known
$g_2(r)$ curves for Laughlin and Moore-Read states.


\section{Calogero-Sutherland model}  \label{sec_CG}

The Calogero-Sutherland model holds a special place in condensed matter theory
as an exactly solvable model which possesses fractional excitations
\cite{CalogeroSutherland}.  The model has Jastrow-type ground state
wavefunctions similar to FQH wavefunctions; the eigenfunctions are known in
terms of Jack polynomials.  Using
properties of Jack polynomials (`duality relations'),
Ref.~\cite{KatsuraHatsuda_CalogeroSutherland_JPA07} has studied particle
partitioning entanglement entropies for this model.  

The model is described by the Hamiltonian ($0\leq{x}\leq{L}$):
\[
 \hat{H} ~=~ - \frac{1}{2}\sum_{i=1}^{N} \frac{\partial^2}{\partial x_i^2} 
~+~ \sum_{i<j} 
 \frac{\beta(\beta-1) \lba\frac{\pi}{L}\rba^2}{\sin^2 \lba \frac{\pi}{l} (x_i - x_j) \rba}
 \; .
\]
%
%
By utilizing properties of Jack polynomials in the limit $N-n \rightarrow
\infty$, Ref.~\cite{KatsuraHatsuda_CalogeroSutherland_JPA07} provides the
following upper bound on the $n$-particle entanglement entropy $S_n$ in the
$N$-particle Calogero-Sutherland ground state:
\[
 S_{n} \leq S_{n}^{bound} = \ln \lba  \begin{array}{c} \beta(N-n) + n \\ n 
 \end{array}
 \rba  \; .
\] 
In close analogy to the FQH results discussed in Section \ref{sec_FQH}, this
expression can be interpreted in terms of exclusion statistics.  Tracing out
one particle is equivalent to removing one particle from 
Fermi sea, which in turns creates $\beta$ quasiholes. Thus taking out $N-n$
particles creates $\beta (N-n)$ quasiholes.  There are $C(\beta(N-n)+n, n)$
ways to accommodate $n$ particles and $\beta (N-n)$ quasiholes within the
Fermi sea, which gives the above estimate.

Ref.~\cite{KatsuraHatsuda_CalogeroSutherland_JPA07} also gives an explicit
expression for the sub-leading correction to this bound:
\[
  S_{N} ~\simeq~ \ln \lba \frac{(\beta(N-n) + n)!}{n! (\beta(N-n))!}\rba ~-~ 
n (\ln\beta -1 + \beta^{-1}) ~+~ \mathcal{O}(N^{-1})  \; . 
\]
In this formula the first term originates from the number of nonzero
eigenvalues whereas the second one comes from the asymptotic eigenvalue
distribution.  An explicit expression for the $\mathcal{O}(N^{0})$ term is
possible because of the more detailed understanding of the eigenvalue
distribution (in terms of the relevant Young tableaus) that is available for
this model, as compared to the FQH case.

\vspace{1cm}

\section{Conclusions}

Particle entanglement is an emerging important measure of correlations in
itinerant many-particle quantum systems.  In this review, we have provided an
extremely simple explicit example (section \ref{sec_elementary-example})
designed to clarify the concept and remove misconceptions.  We have
synthesized the available results into a set of general observations (section
\ref{sec_general}).  
We have pointed out several different mechanisms for particle entanglement in
itinerant quantum states, such as absent or imperfect Bose condensation,
anti-symmetrization of fermionic systems, and Schr\"odinger cat states.  

The bulk of this review (section \ref{sec_bosons} onwards), of course, surveys
the available results on particle entanglement in itinerant many-body systems.
The discussion of more conventional bosonic, fermionic and even anyonic
systems, falls mostly within the general observations of Section
\ref{sec_general}, with more subtle effects (\emph{e.g.}, anyonic statistics)
showing up as sub-leading terms.  The study of particle partitioning in quantum
Hall states and the Calogero-Sutherland model provides a window into more
exotic phenomena, such as exclusion statistics.

Clearly, the study of particle partitioning entanglement is at its infancy,
and one expects further insights and new calculations to appear.  We hope this
review will provide inspiration for further advances.

\ack

Raoul Santachiara provided data for Figure \ref{fig_anyons}.  During the
course of our earlier work on particle-partitioning entanglement, we received
helpful comments from P.~Calabrese, J.-S.~Caux, J.~Moore, R.~Santachiara, and
J.~Vidal.  
O.S.Z.\ and K.S.\ are supported by the Stichting voor Fundamenteel
Onderzoek der Materie (FOM) of the Netherlands.
M.H.\ acknowledges funding from the ESF (INSTANS programme) for a
collaboration visit to complete this review.

\end{document}